\documentstyle[11pt]{article}
\begin{document}
\centerline{ Ground State Spin Oscillations of a Two-Electron Quantum Dot}
\vskip 0.1cm
\centerline{ in a Magnetic Field}
\vskip 0.3cm
\centerline{ M. Dineykhan$^{1}$ and R.G. Nazmitdinov$^{1,2}$}
\centerline{$^{1}$Bogoliubov Laboratory of Theoretical Physics}
\centerline{Joint Institute for Nuclear Research, 141980 Dubna, Russia}
\centerline{$^{2}$ Centre for Nonlinear Science and Department of 
Physics}
\centerline{University of the Witwatersrand, P.O. Wits 2050, 
Johannesburg, South Africa}
\vskip 0.5cm

\begin{abstract}
Crossings between spin-singlet and spin-triplet
lowest states are analyzed within the model of a two-electron
quantum dot in a perpendicular magnetic field. The explicit
expressions in terms of the magnetic field, the magnetic quantum
number $m$ of the state and the dimensionless dot size
for these crossings are found.
\end{abstract}

\vskip 0.2cm

PACS number(s):73.20Dx, 73.23Ps

\vskip 0.3cm

Recent progress in nanostructure technology allows
to thoroughly investigate  a quantum motion of confined electrons
in semiconductors, in particular, in quantum dots \cite{TC,Kas}.
Since the size of a quantum dot as well as the number of electrons 
are controlled, we can study different correlation effects in 
the "artificial atom". Two interacting electrons in an external 
potential turn out to be a useful system 
providing a basis for understanding the contribution of
different components of an effective mean field.
A detailed examination of the electron 
spectrum of a quantum dot can also be approached
with alteration of the magnetic field \cite{As}.

Low-lying energy levels of a two-dimensional two-electron
system in a perpendicular magnetic field have been analyzed
quantitatively for various confining potentials
in \cite{Wag,Pf,Peet}. In particular,
for a parabolic confining potential
the estimations for singlet-triplet and triplet-triplet
ground state phase transitions have been found in the perturbation 
approach in the limit of the strong magnetic field  \cite{Wag}.
The energy spectrum of two interacting electrons in the parabolic 
confining potential has been obtained analytically for particular values 
of the magnetic field in \cite{Tau}. In this Letter we present 
the analytical expressions for fullfilment of 
such phase transitions at arbitrary values of the magnetic 
field for a parabolic quantum dot.

Our analysis is based on the oscillator representation method (ORM) developed
in \cite{Din}. The application of this method to calculate
the energy spectrum of the two-electron system in the magnetic field
is discussed in \cite{DN} where the reader can find all necessary details.
We consider the two-dimensional version of the model
Hamiltonian \cite{DN}
\begin{eqnarray}
\label{ham}
H& = &\sum_{j=1}^2{\Bigg\{} {1\over 2  m^{\star}}
(\vec  p_j -{e\over c}\vec  A_j)^2+{m^{\star}\over 2}\left[
\omega _0^2(x_{j}^2+y_{j}^2)+\omega _z^2z_{j}^2 \right] {\Bigg\}}\nonumber\\
     \\
& + &\frac{e^2}{4\pi \epsilon \epsilon_0} \frac{1}{|\vec r_1 - \vec r_2|}
+ H_{spin}~,\nonumber
\end{eqnarray}
assuming $z=0$. Here $H_{spin} = g(\vec s_1 + \vec s_2) \vec B$
and  $m^\star$ is the
effective electron mass. Below, we use the units $(e=c=1)$.
For the perpendicular magnetic field
$(\vec B||z)$ we choose a gauge described by the
vector $\vec A =[\vec B\times  \vec r]/2  ={1\over2}\vec B (-y,x,0)$.
By introducing the relative and center-of-mass coordinates
$\vec r = \vec r_2 - \vec r_1$, $ \vec R = {1\over2}(\vec r_1
+\vec r_2)$, the Hamiltonian, Eq.(\ref{ham}), can be separated into the
center-of-mass (CM)  and relative-motion (RM) terms
\begin{eqnarray}
\label{gen}
H& = &2H_q + {1\over 2}H_Q + H_{spin}\\
\label{rel}
H_q& = &{1\over 2}\left[ {\vec p_q} + {\vec A}_q \right]^2 +
{\hbar^2\over 2} {\omega_q}^2 \rho_q^2 +
\frac{k \sqrt{\hbar \omega_0}}{2\rho_q} \\
\label{cen}
H_Q& = &{1\over2}\left[ {\vec P_Q} + {\vec A}_Q\right]^2 +
{\hbar^2\over 2}{\omega_Q}^2 \rho_Q^2
\end{eqnarray}
where $\omega_Q = 2\omega_0$,
$\omega_q = {1\over 2}\omega_0$,
${\vec A}_Q = {\vec A}(q_1) + {\vec A}(q_2)$,
${\vec A}_q = {1\over 2}\left( {\vec A}(q_2) - {\vec A}(q_1)\right)$
and ${\vec A}(q) =\frac{\hbar}{m^\star}[\vec B\times  \vec q]$.
Here we use the variables
$\vec q = \frac{\sqrt{m^\star}}{\hbar}\vec r$,
$\vec Q = \frac{\sqrt{m^\star}}{\hbar}\vec R$,
 $\rho_q = \sqrt{q_x^2+q_y^2}$ and define
the characteristic lengths: the effective radius
$a^{\star} = a_B \epsilon\frac{ m_e}{m^\star}
 (a_B = 4\pi \epsilon_0 \frac{{\hbar}^2}{m_e e^2})$ and
the oscillator length $l_0 =  (\hbar/m^\star \omega_0)^{1/2}$.
These units allow one to define the dimensionless dot size
$k = l_0/a^\star$ \cite{Wag}.   At  $k=0$, we have the model of 
noninteracting electrons in the magnetic field \cite{Fock} 
(see also \cite{As}).
The separability and the conservation of the angular momentum
lead to a natural ansatz for the eigenfunction of the
Hamiltonian, Eq.(\ref{gen}), i.e.
$\Psi = \psi(\vec q)\phi(\vec Q)\chi(\vec s_1, \vec  s_2)$.
According to the Pauli
principle, if the spatial part of the total wave function is
symmetric (antisymmetric) with respect to the inversion
$r\rightarrow -r$, $\chi$ must be a singlet (triplet)
spin state.
For the eigenvalues we have
\begin{equation}
\label{sum}
E=2\epsilon_r + {1\over 2}E_{N,M}+E_{S}.
\end{equation}
Here $\epsilon_r$ and $E_{N,M}$ are the eigenvalues of the
Hamiltonian of the RM and of the CM motion, respectively.
The solution to the CM Hamiltonian $H_Q$ is 
well known \cite{Fock} and the energy can be written as
\begin{equation}
\label{ec}
E_{N,M} =2\hbar {{\omega}_0} \left[ (2N + |M| + 1)\sqrt{1+{t^2\over 4}}
 + {1\over 2}Mt\right]
\end{equation}
where $N=0,1,...$ and $M=0,\pm 1,...$ are radial and azimuthal
quantum numbers, respectively,
and $t = \omega_c/{\omega_0}$ where
$\omega_c=\frac{B}{m^\star}$ is the cyclotron frequency.
The spin of the two electrons leads to the additional Zeeman energy
\begin{equation}
\label{es}
E_S= g^\star \mu_B B S_z =
\frac{\hbar {{\omega}_0}}{4}\left[ 1-(-1)^m \right] g^\star
\frac{m^\star}{m_e}t
\end{equation}
Here, $m$ is a magnetic quantum number corresponding to 
the RM Hamiltonian
and $g^\star$ is the effective Lande factor.
For the RM energy we obtain \cite{eq}
\begin{equation}
\label{er}
E_{nm}^{2d} =  \epsilon_r = {\varepsilon}_{nm}^0 +
{\varepsilon}_{nm}^c
\end{equation}
where
\begin{eqnarray}
\label{er1}
{\varepsilon}_{nm}^0 & = & \frac{\hbar {\omega}_0}{2}
\left[ {m\over 2}t +
(1+|m|+2n)x^2(1+\frac{t^2}4)^{1/2}\right]\nonumber\\
   \\
{\varepsilon}_{nm}^c & = &
\frac{\hbar {\omega}_0}{2} \frac{x k}{2\sqrt{2}}
(1+\frac{t^2}4)^{1/4}\left[ 3
\frac{\Gamma({1\over 2}+|m|)}{\Gamma(1+|m|)}+2<n|h_I|n>\right]\nonumber
\end{eqnarray}
The quantity $x$ is determined by the following equation:
\begin{equation}
\label{x}
x^4 + \frac{x^3}
{\sqrt 2} \frac{k}{(1+\frac{t^2}4)^{1/4}}
\frac{\Gamma({1\over 2}+|m|)}{\Gamma(2+|m|)}
-1 = 0.
\end{equation}
In the ORM only the lowest positive solution 
of Eq.(\ref{x}) is interesting.
In contrast to the perturbative approach 
the main term ${\varepsilon}_{nm}^0 $, Eq.(\ref{er1}),
depends on the Coulomb forces as well. The interaction modifies the external
potential and results in the effective mean field potential for
the RM.
Notice that in the second term ${\varepsilon}_{nm}^c$
there is a contribution arising due to
radial excitations, i.e., the term $<n|h_I|n>$ (see \cite{DN}).
The matrix element $<n|h_I|n>$
provides a basis for calculations of the radial excitations modified by
the Coulomb interaction and, according to the rules of the ORM,
it has the following form:
\begin{eqnarray}
\label{5,2.19}
<n|h_I|n>= \frac{3}{4}\cdot
\frac{\Gamma(d/2-1/2)}{\Gamma(d/2)}\cdot S_{n}
\end{eqnarray}
where
\begin{eqnarray}
\label{5,2.20}
S_n=\frac{4\Gamma(1+n) }{3\sqrt{\pi}}\cdot\sum_{k=2}^{2n}
\frac{(-1)^k\Gamma(k+1/2)}{\Gamma(k+d/2)}\cdot N_k(n,d)~,\nonumber\\
N_k(n,d)=\sum_{p=0}^{n}\frac{2^{2p-k}\Gamma(k+n-p+d/2)}
{(n-p)!(2p-k)!\left((k-p)!\right)^2}~.
\nonumber
\end{eqnarray}
and $d=2+2|m|$.
For particular values of the radial quantum number $n=0,1,2$
we obtain $S_0=0$,
$ S_1=\frac{2}{d}~$, $
S_2=\frac{4}{d(d+2)}\cdot\left[d+\frac{19}{8}\right]~$,
respectively. Our result for the RM energy,
Eqs.(\ref{er}) and (\ref{er1}), corresponds
to the perturbation approach in the limit $x\rightarrow 1$,
$\omega_c >> \omega_0$ and $n=0$ (see below).

Owing to the separability of the CM energy and the RM energy,
we have only two frequencies from Eq.(\ref{ec}) for dipole transitions
$(N M)\rightarrow (N^{'}M^{'})$
\begin{equation}
\omega_{\pm}=[(\frac{\omega_c}2)^2+\omega_0^2]^{1/2} \pm \frac{\omega_c}2
\end{equation}
which are not influenced by the Coulomb interaction.
This is a simple example of the consequence of the Kohn theorem
\cite{Jon}. Since the center-of-mass quantum numbers
$N, M$ and the quantum number $m$ are conserved by the Coulomb interaction,
the ground state has the quantum numbers $N=0$, $M=0$, $n=0$. Comparing
the energy with different $m\leq0$, we can define the ground state energy
at different values of the magnetic field ${\omega_c}/{{\omega}_0}$.
While without the Coulomb forces the ground state is
always the state with $m=0$, the Coulomb interaction leads
to a sequence of different ground states $m= -1, -2,...$ which are
an alternating sequence of singlet and triplet states.
In Fig.1 we have plotted the energy of states with different $m$
for different values of $k=l_0/a^*$. 

The singlet-triplet ground phase transition (crossing)
occurs when the following condition is fulfilled
$E_{0,m} = E_{0,m-1} (m\leq0)$ where we have introduced the notation
$E_{n,m}=E_S+2E_{n,m}^{2d}$. Taking into account Eqs.({\ref{es}),
(\ref{er}), we obtain
\begin{eqnarray}
\label{x1}
\frac{\ell_0}{a^*}{\Big|}_{n=0}&=&\frac{2\sqrt{2}}{3x}
\cdot\frac{\Gamma(2+|m|)}{\Gamma(1/2+|m|)}
\cdot\frac{t}{\left[1+t^2/4\right]^{1/4}}
\nonumber\\
&\times&\left[-1+
x^2\sqrt{1+\frac{4}{t^2}}
+(-1)^mg^*\frac{m^*}{m_e}
\right]~.
\end{eqnarray}
We emphasize that this expression is valid at finite values of
the magnetic field.
Combining Eq.(\ref{x}) with Eq.(\ref{x1}), we can define 
the value of $x$ at the singlet-triplet crossings $m\rightarrow m-1$
\begin{eqnarray}
\label{x2}
x=\sqrt{\frac{\sigma t
+\sqrt{\left(\sigma^2+\frac{21}{4}\right)t^2+21}}
{7\sqrt{1+\frac{1}{4}t^2}}}
\end{eqnarray}
where
\begin{eqnarray}
\label{bet}
\sigma=1-(-1)^mg^*\frac{m^*}{m_e}=1-2\beta~,~~~~
\beta=\frac{1}{2}(-1)^mg^*\frac{m^*}{m_e}~.
\end{eqnarray}
Therefore, Eqs.(\ref{x1}) and (\ref{x2})  define the values
of the magnetic field which lead to the singlet-triplet crossings
in the ground state and in the excited states of the two-electron
system as well. At the strong magnetic field $(\omega_c >> \omega_0)$
in the lowest order of the parameter $\beta$
we obtain from Eqs.(\ref{x1}) and (\ref{x2})
\begin{equation}
\label{xx}
x\simeq 1-\frac{1}{5} (-1)^m g^* \frac{m^*}{m_e}
\end{equation}
\begin{eqnarray}
\label{x3}
\frac{\ell_0}{a^*}{\Big|}_{n=0}&=&\frac{8}{3}
\frac{\Gamma(2+|m|)}{\Gamma(1/2+|m|)}\nonumber\\
&\times&\left[\left(\frac{\omega_0}{\omega_c}\right)^{3/2}+
\frac{3}{10}(-1)^mg^*\frac{m^*}{m_e}\left(
\frac{\omega_c}{\omega_0}\right)^{1/2}\right]
\end{eqnarray}
Eq.(\ref{x3}) almost coincides with
a similar expression based on the perturbative approach \cite{Wag} 
with the exclusion of the coefficient in front of the term arising 
from the spin contribution.

For the negative Lande factor the spin-splitting energy in
the magnetic field will lower the energy of the spin $S_z = +1$
component of the triplet states. In particular, the relation
$E_{0, m}=E_{0, m-1}= E_{0, m-2}$ (m odd) defines the point when
the singlet phase ceases to exist \cite{Wag}. Beyond this point
we can observe crossings  between triplet states
defined by the condition
$E_{0,m} = E_{0,m-2}$ (m odd), which leads to
\begin{eqnarray}
\label{x4}
\frac{\ell_0}{a^*}&=&\frac{8\sqrt{2}}{3x}
\frac{t}{\left[1+t^2/4\right]^{1/4}}\cdot
\frac{\Gamma(3+|m|)}{\Gamma(1/2+|m|)}\\
&\times&\frac{1}{5+4|m|}\cdot\left(-1+x^2\sqrt{1+\frac{4}{t^2}}\right)
\nonumber
\end{eqnarray}
For the triplet-triplet crossings the parameter $x$
can be obtained from Eqs.(\ref{x1}) and (\ref{x4})
\begin{eqnarray}
\label{x5}
x^2&=&\frac{4t}{\sqrt{1+t^2/4}}\cdot\frac{2+|m|}{47+28|m|}\\
&+&\sqrt{\frac{16t^2}{1+t^2/4}\cdot
\frac{(2+|m|)^2}{(47+28|m|)^2}
+\frac{3(5+4|m|)}{47+28|m|}}
\nonumber
\end{eqnarray}
At the strong magnetic field $\omega_c >>\omega_0$
for the triplet-triplet crossings $m\rightarrow m-2$ (m odd)
we obtain the following result:
\begin{eqnarray}
\label{x6}
x& \simeq & 1~,\\
\frac{\ell_0}{a^*}&=&\frac{8}{3}
\frac{\Gamma(3+|m|)}{\Gamma(1/2+|m|)}
\frac{1}{5+4|m|}\cdot\left(\frac{\omega_0}{\omega_c}\right)^{3/2}~,
\nonumber
\end{eqnarray}
which coincides with the estimation of the perturbative
approach \cite{Wag}. The singlet-triplet crossings
yield the triplet-triplet ones for
polarized electrons. The ground states
are determined with odd values of the magnetic quantum number
$m$. The radius of each particular $m$ state decreases as
$1/\sqrt{B}$, and the electrons are pushed towards the
dot's center. At very high fields both the electrons are in 
the lowest Landau level \cite{Lan} which, however, consists of
the quantum levels, Eqs.(\ref{er}) and (\ref{er1}). Naturally,
the Pauli principle prevents the occupation of
the same quantum state by both the electrons.
The Coulomb forces become less important
for high-lying single-electron levels. The orbital motion
increases the relative distance between electrons weakening
the influence of the Coulomb forces on the crossing of levels.
The value of the parameter $k=l_0 /a^{\star}$ for
the singlet-triplet crossing decreases
with increasing the radial quantum number $n$ as well.
In particular, for the two-dimensional system we have obtained the
following relation between the
parameters $k=l_0 /a^{\star}$ for the singlet-triplet crossing at
different $n$
\begin{equation}
\frac{(l_0 /a^{\star})_{n=1}}{(l_0 /a^{\star})_{n=0}} = \frac{2+|m|}{7+|m|}
\end{equation}
While  the interplay between the magnetic field
and the Coulomb forces determines the features of the phase transition
(singlet $\rightarrow$ triplet) for the ground state ($n=0$),
mainly the magnetic field
leads to phase transitions for the high-lying states $n>0$.

In conclusion, we have found the exact relations between
the dimensionless size $k=l_0/a^*$ and the values of
the magnetic field for the singlet-triplet and
the triplet-triplet crossings in the ground and excited states
for two interacting electrons confined by the  parabolic potential.
The Coulomb interaction is treated exactly within the analytical approach.
The spin interaction is important for the singlet-triplet
crossings whereas it does not contribute to the triplet-triplet
crossings. The last one can be described as the crossing of quantum
levels of polarized electrons in the magnetic field.

We are grateful to  Viktor Kabanov for useful discussions.
R.G.N. acknowledges financial support from the Foundation for Research
Development of South Africa which was provided under the auspices of the
Russian/South African Agreement on Science and Technology.

Figure Caption

{\bf Fig.1} Eigenenergies in units $\hbar \omega_0$ vs
the ratio $\omega_c/\omega_0$ for a different dot size $k$.
The family of states $N=M=n=0$ and $m\leq0$ is
shown (a)for $ k=2$ and (b) for $k=4$. As the ratio $k=l_0/a^*$
increases the Coulomb interaction rearranges the 
sequence of levels.


\begin{thebibliography}{99}
\bibitem{eq} we corrected the misprint, i.e. the factor $2$,
which presents in the equation for the energy of relative motion
in \cite{DN}, is omitted here.
\end{thebibliography}
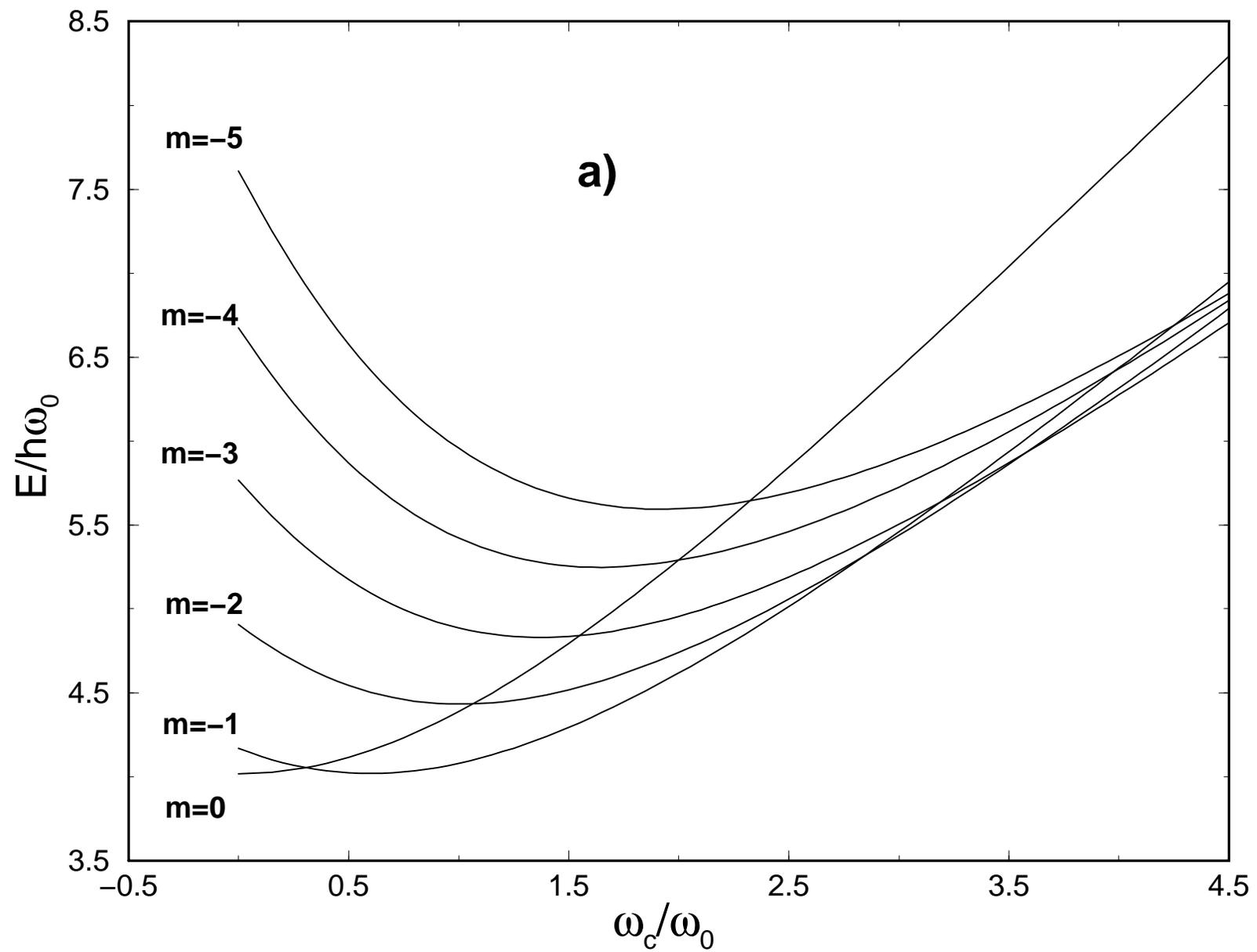
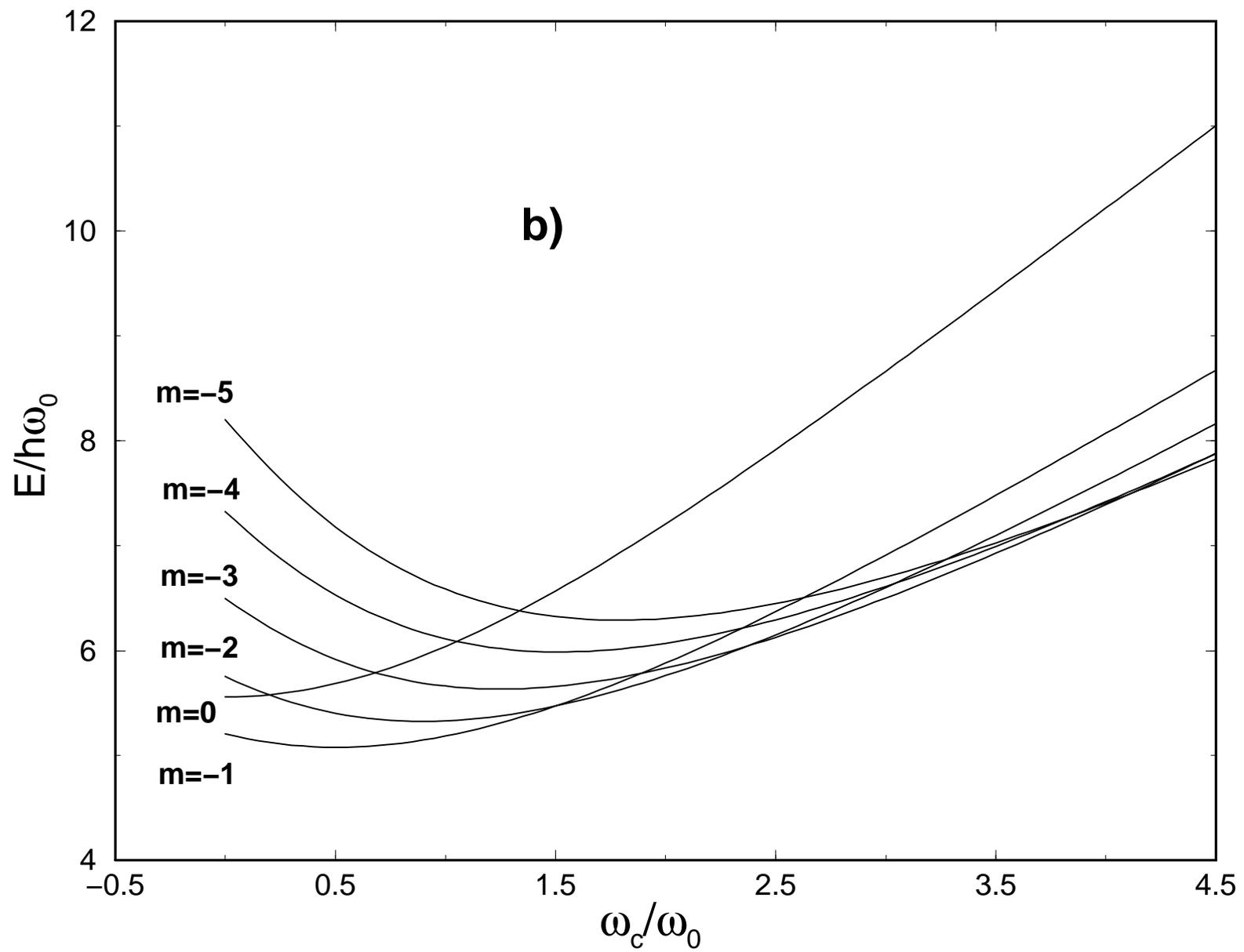
\end{document}